\documentclass{appolb}
\usepackage{epsfig}
\newcommand{\lsim}{\raisebox{-0.13cm}{~\shortstack{$<$ \\[-0.07cm] $\sim$}}~}
\newcommand{\gsim}{\raisebox{-0.13cm}{~\shortstack{$>$ \\[-0.07cm] $\sim$}}~}
\begin{document}
\title{EXOTIC PHYSICS WITH ULTRAHIGH ENERGY COSMIC RAYS%
\thanks{Presented at the XXXI International Conference of Theoretical Physics, 
        Matter To The Deepest: Recent Developments In Physics of Fundamental Interactions,
        5-11 September 2007, Ustro\'n, Poland.}%
}
\author{Jos\'e I. Illana$^{\rm a}$,
\address{$^{\rm a}$CAFPE and Departamento de F{\'\i}sica Te\'orica y del Cosmos, \\
                   Universidad de Granada, E-18071 Granada, Spain}
\and
Markus Ahlers$^{\rm b}$, Manuel Masip$^{\rm a}$, Davide Meloni$^{\rm c}$
\address{$^{\rm b}$Deutsches Elektronen-Synchrotron DESY,
                   D-22607 Hamburg, Germany \\
         $^{\rm c}$Dipartimento di Fisica, Universit\'a di Roma Tre, 
                   I-00146 Roma, Italy}
        }
\maketitle
\begin{abstract}
Ultrahigh energy cosmic rays provide a unique ground for probing new physics.
In this talk we review the possibility of testing TeV gravity in interactions of cosmogenic neutrinos and the potential to discover long-lived exotic particles in nucleon-produced air showers, such as gluinos of split-SUSY models or staus of supersymmetric models with a gravitino LSP.
\end{abstract}

\vskip-11cm
\hskip 7cm CAFPE-89/07, UG-FT-219/07
\vskip11cm

\section{Introduction}

Cosmic rays reach the Earth with energies several orders of magnitude above those achieved in colliders. To probe this region of ultrahigh energy (UHE), a new generation of experiments is already operating or being deployed. In particular, the Pierre Auger Observatory in Argentina for the detection of extensive air showers and the neutrino telescope IceCube at the South Pole. They have the potential to find new physics. 

\section{TeV gravity explored by cosmogenic neutrinos}

The cosmogenic neutrinos are produced in the scattering of protons off cosmic microwave background photons. Their flux, yet unobserved, depends on the production rate of primary nucleons of energy around and above the GZK cutoff. It is correlated with proton and photon fluxes that must be consistent, respectively, with the number of ultrahigh energy events at AGASA and HiRes and with the diffuse $\gamma$-ray background measured by EGRET. 
A {\em higher} ({\em lower}) neutrino flux assumes that the photon flux saturates (accounts for 20\%) of EGRET observations \cite{semikoz}.
The spectrum has a peak at neutrino energies between $10^9$~GeV and $10^{10}$~GeV.

These neutrinos have access to TeV physics in interactions with terrestrial nucleons at center of mass energies $\sqrt{s}=\sqrt{2m_NE_\nu}\gsim10$~TeV. If the fundamental scale of gravity is $M_D\sim1$~TeV \cite{ADD}, which may happen in $D>4$ spacetime dimensions, these processes are transplanckian.
The only consistent theory known so far in such a regime, string theory, tells us that the interactions are soft in the ultraviolet. The scattering amplitudes vanish except in the forward region, an effect that can be understood as the destructive interference of string excitations \cite{SR}. The forward amplitudes are dominated by the zero mode of the string, corresponding to the exchange of a gauge particle of spin 1, ${A}\sim gs/t$, for open strings, or a graviton of spin 2, ${A}\sim (1/M^2_D) s^2/t$, for closed strings. Therefore, one expects that gravity dominates in transplanckian collisions. Furthermore, such collisions probe $M_D$ directly and independently of compactification details.
In gravitational interactions one must keep in mind two critical values in the impact parameter space: the Planck length $\lambda_D\sim M_D^{-1}$ and the Schwarzschild radius $R_S(s)\sim(\sqrt{s}/M_D)^{1/(n+1)}M_D^{-1}$. There are two types of interactions \cite{riccardo}:

Short-distance interactions, with impact parameter $b\lsim R_S$, in which the colliding particles (a neutrino and a parton inside the nucleon) collapse into a black hole (BH). The geometric cross section $\hat\sigma_{\rm BH}\simeq\pi R_S^2(\hat s)$ for the partonic process, with $\hat s=xs$, is a good approximation for $\hat s\gg M_D^2$ \cite{BH}.  However, most of the BHs produced in the scattering of an ultrahigh energy neutrino off a parton are light, with masses just above $M_D$, which adds uncertainty to this estimate.

Long-distance interactions, with $b\gg R_S$, are due to the exchange of weakly coupled gravitons of low momentum (linearized gravity) \cite{eik}. In transplanckian collisions quantum gravity acts at distances of order $\lambda_D$, well within the event horizon ($R_S>\lambda_D$). These interactions are characterized by a small deflection angle in the center of mass frame. The elastic collision of a neutrino and a parton that exchange $D$-dimensional gravitons is described by the {\em eikonal} amplitude resumming an infinite set of ladder and cross-ladder diagrams. In these {\em soft} processes, the neutrino transfers to the partons a small energy fraction $y=(E_\nu-E'_\nu)/E_\nu$ and keeps going. Then the parton starts a hadronic cascade, that will be observable if $y>E_{\rm thres}/E_\nu$.

To estimate the relative frequency of both type of processes \cite{TG}, consider a  $10^{10}$~GeV neutrino that scatters off a nucleon with $E_{\rm thres}=100$~TeV and $M_D=1$~TeV for $n=2\;(6)$ extra dimensions. The number of eikonal interactions before the neutrino gets destroyed is the ratio of interaction lengths $L_{\rm BH}/L_{\rm eik}=\sigma_{\rm eik}/\sigma_{\rm BH}\approx 14\;(1.6)$, with $L_{\rm BH}\approx17$~km (4 km) in ice, whereas for a SM interaction $L_{\rm SM}\approx440$~km. This opens up the possibility of {\em multiple-bang} events. In fact, for a detector of length $L$, the total interaction probability, the probability of $N$ bangs and the average number of bangs is
$$
P_{\rm int}(L)=\sum_{N=1}^\infty P_N(L), \
P_N(L)={\rm e}^{-L/L_\sigma}\frac{(L/L_\sigma)^N}{N!},\ 
\langle N\rangle=\sum_{N=1}^\infty NP_N=\frac{L}{L_\sigma},
$$
which yields, for the given neutrino and $L=1$~km, $P^{\rm SM}_{\rm int}=0.0022$, $P^{\rm BH}_{\rm int}=0.06\ (0.22)$, $P^{\rm eik}_1=0.36\ (0.27)$, $P_2^{\rm eik}=0.15\ (0.06)$ and
$P^{\rm eik}_{>2}=0.05\ (0.008)$.

\begin{figure*}[t]
\begin{minipage}[t]{0.48\linewidth}
\includegraphics[width=\linewidth]{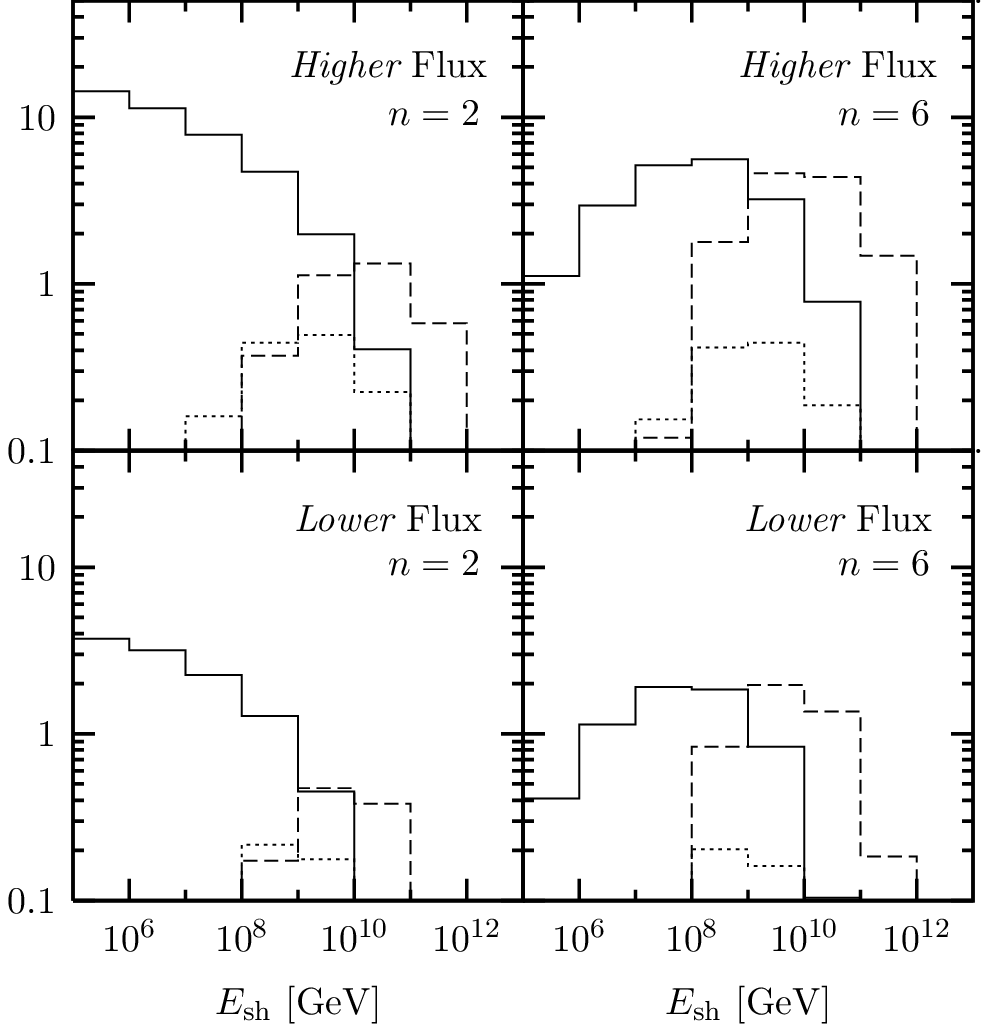}
\caption{
Energy distribution (events per bin) of the eikonal (solid), BH (dashed)
and SM (dotted) events in IceCube per year for the {\it higher} and
the {\it lower}
cosmogenic fluxes, $M_D=2$~TeV and $n=2,\ 6$.}
\label{fig1}
\end{minipage}\hfill
\begin{minipage}[t]{0.48\linewidth}
\includegraphics[width=\linewidth]{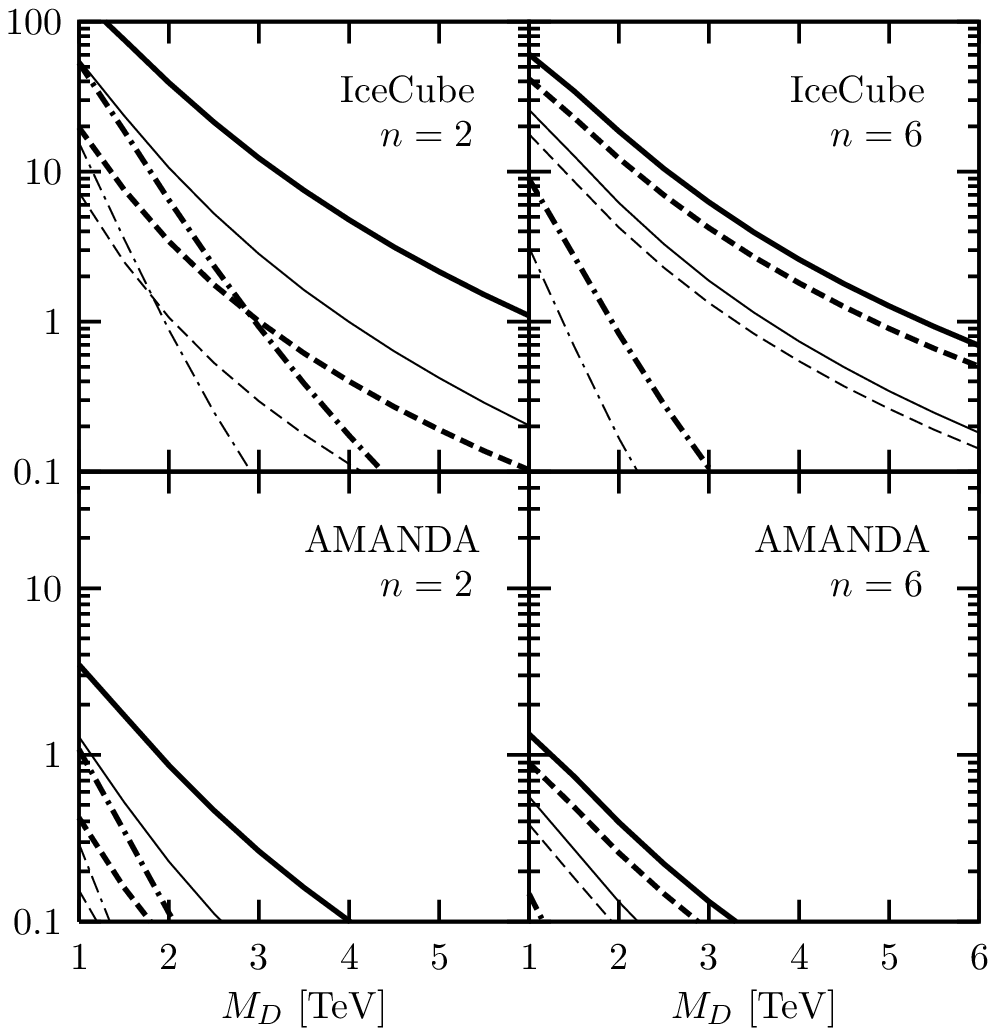}
\caption{
Contained events per year in IceCube and AMANDA for
{\it higher} (thick) and {\it lower} (thin) fluxes,
showing eikonal (solid), multi-bang (dashed-dotted) and BH (dashed)
events.}
\label{fig2}
\end{minipage}
\end{figure*}

The event rate at a neutrino telescope is proportional to the flux, the exposure time, the cross sectional area, the interaction probability (roughly proportional to the length) and the probability that the neutrino of a given inclination survives to reach the detector (depending on the zenith angle, column density of material and the detector depth).
The energy distribution of the hadronic cascades and the total number of black hole and eikonal events \cite{TG} at AMANDA (0.03 km$^2$ and a length of 700 m) and IceCube (1 km$^3$) for the neutrino fluxes introduced above are given in figs.~\ref{fig1} and \ref{fig2}, respectively. In the SM we expect 1.32 (0.50) contained events per year in IceCube for the higher (lower) flux. Of those, 0.38 (0.14) would come from a neutral current and 0.94 (0.36) from a charged current.

\section{Long-lived exotic particles from UHE nucleons}

The flux of primary nucleons reaching the atmosphere from several GeV follows approximately a power law with negative spectral index 2.7 changing to 3 at about $10^6$~GeV (the {\em knee}) and then back to 2.7 at about $10^{10}$~GeV (the {\em ankle}). Around 90\% of these nucleons are protons and the rest are neutrons, free or bound in nuclei. As this flux enters the atmosphere, an induced flux of secondary hadrons (nucleons, pions, kaons, etc.) is generated \cite{GL}, many of them with enough energy to produce TeV physics in interactions with atmospheric nucleons. For instance, at $10^7$ GeV ($\sqrt{s}\approx5$~TeV) the secondary nucleons (mesons) increase a 50\% (15\%) the number of primaries.

Exotic massive particles \cite{Fairbairn:2006gg} produced inside the shower could be detected at ground if they are {\em long-lived} (their decay products would be otherwise confused with thousands of particles), since they tend to be more {\em penetrating} and can survive the shower (most of ordinary particles, except muons and neutrinos, get aborbed by the atmosphere in sufficiently inclined showers).

\subsection{Gluinos in split-SUSY models}

\begin{figure*}[t]
\begin{minipage}[t]{0.48\linewidth}
\includegraphics[width=\linewidth]{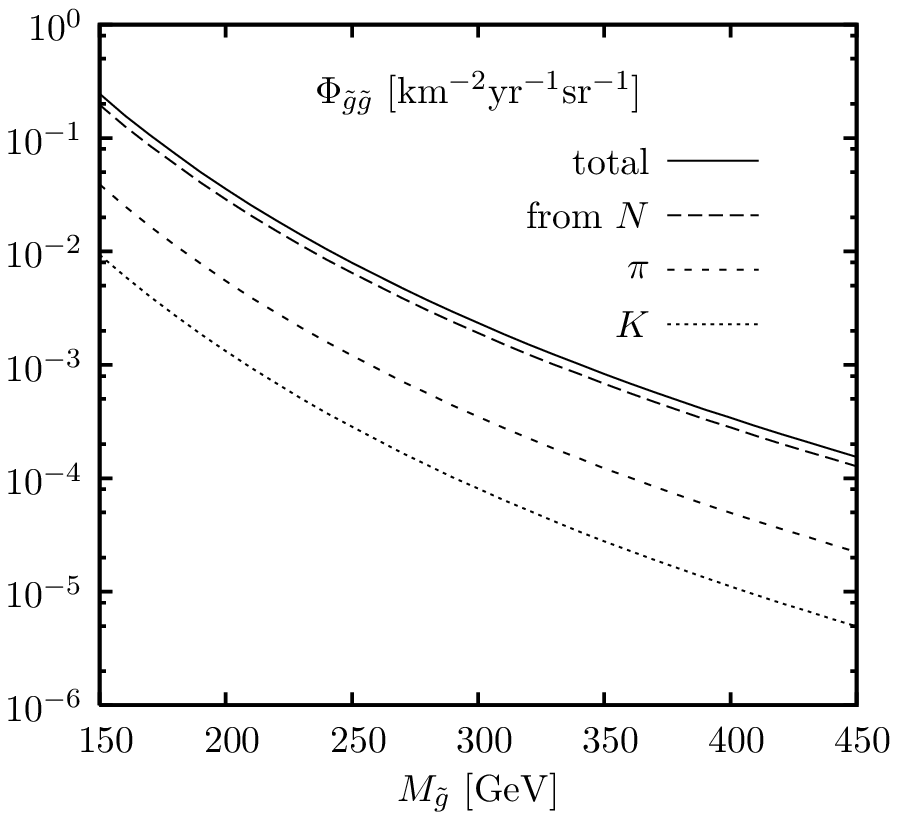}
\caption{
Flux of gluino pairs from different hadrons versus the gluino mass.}
\label{fig3}
\end{minipage}\hfill
\begin{minipage}[t]{0.48\linewidth}
\includegraphics[width=\linewidth]{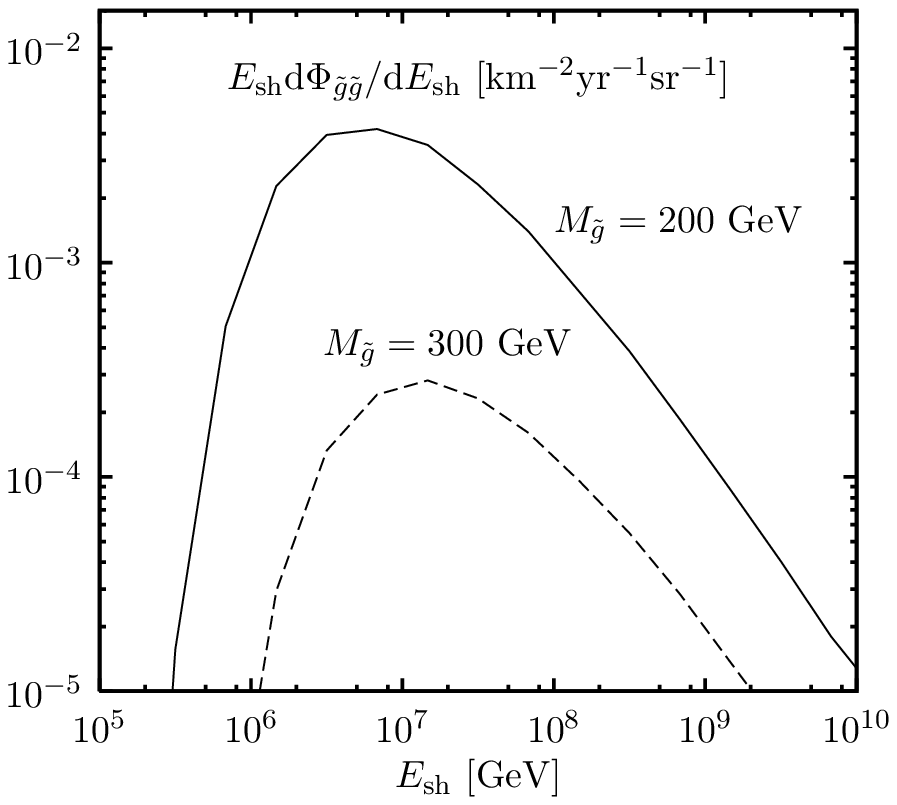}
\caption{
Distribution of gluino pairs as a function of the shower energy.}
\label{fig4}
\end{minipage}
\end{figure*}

In split-SUSY models \cite{Arkani-Hamed:2004fb} the squarks are heavy and the gluinos may be light and long-lived if $R$-parity is conserved. These gluinos would be pair-produced by atmospheric interactions with the UHE hadron flux \cite{GL}, at a rate of less than one down-going pair per year and square kilometer if their mass is just above 160 GeV (fig.~\ref{fig3}). A gluino rapidly fragments into an $R$-hadron. For instance, a neutral gluino-gluon state of mass $M$ (larger than 170 GeV from Tevatron) would have a quite small interaction length (similar to that of a pion at the same velocity) but would be nevertheless very penetrating, losing only $\Delta E/E\approx 0.2 \mbox{ GeV}/M$ per interaction. Thus, while a proton deposits most of its energy in two vertical atmospheres, a 200 GeV gluino gives away just one per cent of its energy.

To observe the long-lived gluinos, first one needs enough events. This will not be the case in a neutrino telescope like IceCube, with an area of 1~km$^2$. In contrast, the Pierre Auger Observatory has an extension of more than 3000 km$^2$ and a shower-energy threshold of about $10^8$~GeV that would allow the detection of 20 (2) gluino-pairs per year if $M=200\ (300)$ GeV (fig.~\ref{fig4}). The gluino signal is distinct, since it consists of a series of hadronic mini-showers resembling a trace of constant energy, more clearly seen in inclined showers (25\% of them have zenith angle above $60^\circ$). Moreover, gluinos come in pairs separated by a distance $D\theta_{\tilde g\tilde g}$, enhanced in quasi-horizontal showers for which $D\approx\sqrt{2HR_\oplus}\approx 250$ m if they start at a typical height $H=20$ km with an average opening angle $\theta_{\tilde g\tilde g}\approx 5\times 10^{-4}$.

\subsection{Staus as long-lived NLSP}

In supersymmetric models with an exact $R$-parity and a gravitino LSP working as dark matter, a charged NLSP, typically a `right-handed' stau, is a long-lived particle since it can only decay via gravitational interactions. To get a sizeable rate the staus must be pair-produced through strong interactions, in prompt decays of coloured particles (gluinos and squarks) \cite{Ahlers:2007js}.

\begin{figure*}[t]
\begin{minipage}[t]{0.48\linewidth}
\includegraphics[width=\linewidth]{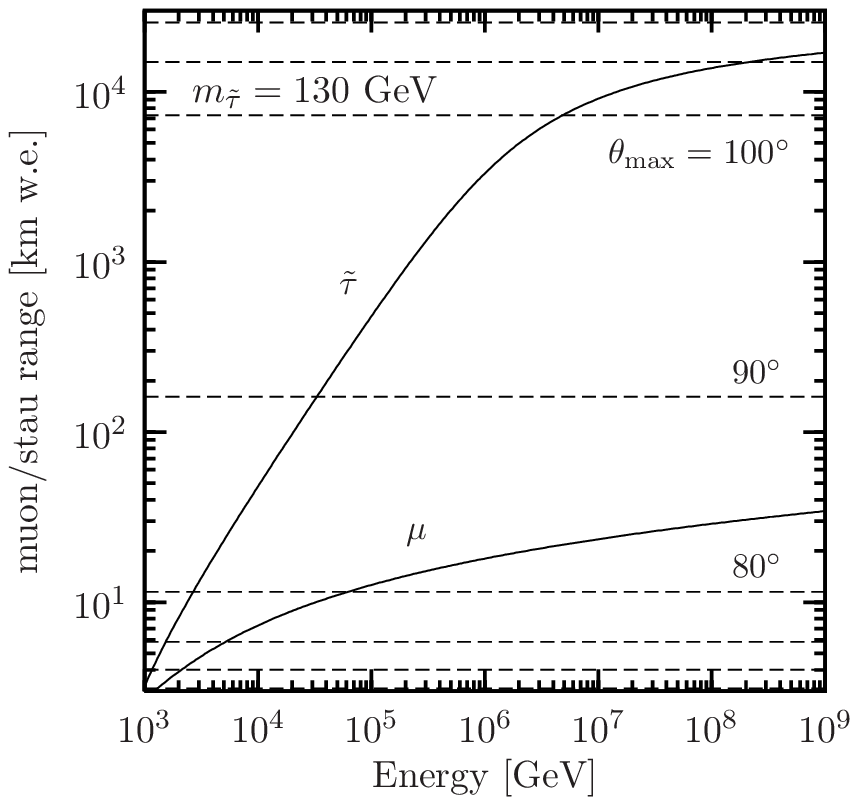}
\caption{
Range of staus and muons versus the energy. The dashed lines show the integrated column-depth from IceCube for different zenith angles.
}
\label{fig5}
\end{minipage}\hfill
\begin{minipage}[t]{0.48\linewidth}
\includegraphics[width=\linewidth]{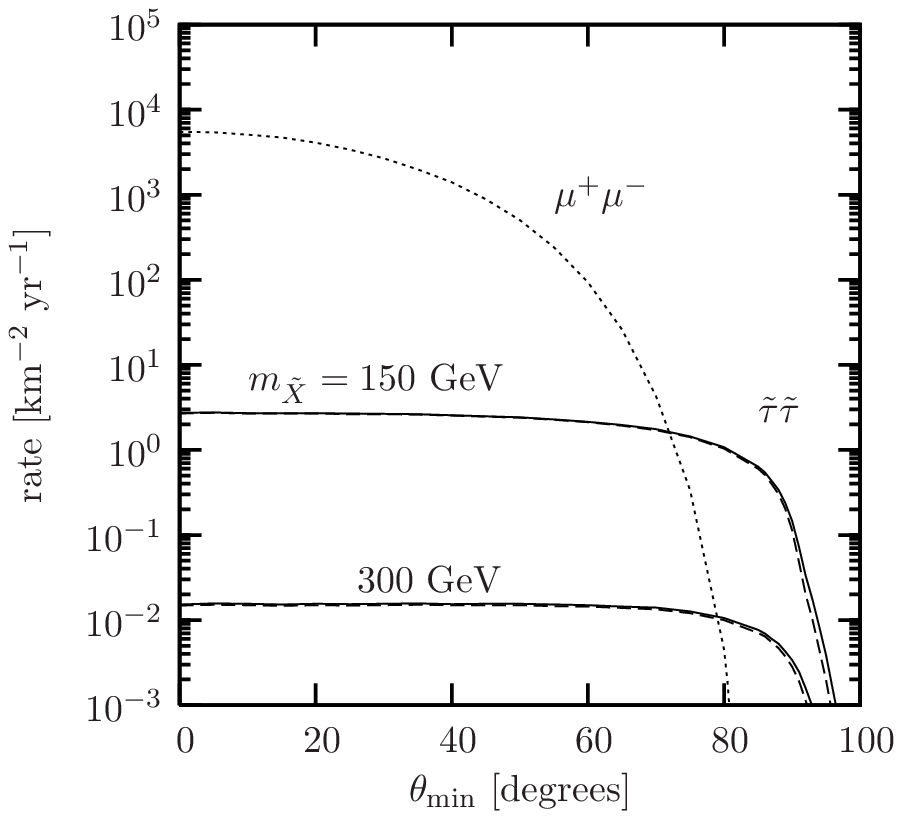}
\caption{
Integrated number of stau and muon pairs separated more than 50 m at IceCube versus the zenith angle.}
\label{fig6}
\end{minipage}
\end{figure*}

The stau-pairs produced high in the atmosphere propagate down to the core of IceCube, about two kilometers below the antarctic ice, facing a strong background of muon pairs. The propagation of muons and charged particles in matter is well understood. The energy deposition per column density follows a law $-{\rm d}E/{\rm d}z=a+bE$ where the first term, independent of the mass of the particle, describes ionization effects and the second, inversely proportional to the mass, accounts for bremsstrahlung, pair production and photohadronic processes. Below 500 GeV ionization is dominant, so staus and muons lose energy at a similar rate, but above that energy the staus, being at least a thousand times heavier, are much more penetrating. In fig.~\ref{fig5} the range of staus and muons is given compared to the integrated column depth of the Earth from IceCube at different zenith angles. For example, a muon of $E=10^7$ GeV has a range of just 25 km water equivalent (w.e.) and can reach IceCube from an angle $\theta_{\rm max}\approx86^\circ$ while a stau of 130~GeV of the same energy can arrive from a $\theta_{\rm max}\approx105^\circ$. Another relevant observable is the separation of the two particles in the detector, which can be resolved if it is larger than 50 m. Fig.~\ref{fig6} shows the number of stau and muon pairs reaching IceCube from a height $H=15$~km as resolved parallel tracks carrying a fraction $\eta=1$ or 0.7 of the energy of parent $\tilde X$ (gluino or squark).

This signal can be distinguished from possible stau pairs produced by cosmogenic neutrinos, extensively considered in the literature \cite{neutrinos}. Firstly, in quasi-horizontal showers the staus produced by neutrinos have a smaller track separation since they are typically produced in the ice, and secondly they may be distributed more isotropic, while those produced by hadrons cannot arrive from much below the horizon.

\section{Conclusions}

The cosmogenic neutrinos directly probe TeV gravity in transplanckian neu\-trino-nucleon collisions, which may be of two types. Hard processes (subdominant) where a microscopic black hole is formed and subsequently evaporates. Soft, elastic processes where a small energy fraction is transfered by the neutrino to a hadronic cascade producing a clear signal in large neutrino telescopes (absence of muons and possibly multiple-bang events).

The primary nucleons or secondary hadrons inside an air shower may produce new massive particles when they interact with atmospheric nucleons. In particular inclined air showers may contain well separated, long-lived gluino pairs (possibly detectable by Auger) or staus
(distinguishable from the muon background close to horizon at IceCube).

JII wishes to thank the organizers of the Conference for their hospitality and the nice atmosphere of the meeting. This work has been supported by MEC of Spain (FPA2006-05294), Junta de Andaluc{\'\i}a (FQM-101 and FQM-437) and a grant CICYT-INFN (07-10).

\end{document}